\documentstyle[11pt,aaspp4]{article}

\slugcomment{}

\lefthead{}
\righthead{}

\begin{document}

\title{The Nature of the Mid-Infrared Background\\
    Radiation in the Galactic Bulge\\
    from the IRTS Observations}

\author{Kin-Wing Chan and T. L. Roellig}
\affil{NASA/Ames Research Center, MS 245--6, Moffett Field, CA 94035; kwc@ssa1.arc.nasa.gov, roellig@ssa1.arc.nasa.gov}

\author{T. Onaka and I. Yamamura\altaffilmark{1}}
\affil{Department of Astronomy, University of Tokyo, Bunkyo-ku, Tokyo 113; onaka@apsun4.astron.s.u-tokyo.ac.jp, yamamura@astron.s.u-tokyo.ac.jp}

\and

\author{T. Tanab\'e}
\affil{Institute of Astronomy, University of Tokyo, Mitaka, Tokyo 181; ttanabe@mtk.ioa.s.u-tokyo.ac.jp}


\altaffiltext{1}{present address:  Astronomical Institute ``Anton Pannekoek'', 
    University of Amsterdam, Kruislaan 403, NL-1098SJ, Amsterdam, the 
    Netherlands}


\begin{abstract}
Using the Mid-Infrared Spectrometer (MIRS) on board the Infrared Telescope in Space (IRTS)
we obtained the 4.5 to 11.7 $\mu$m spectra of the stellar populations and 
diffuse interstellar medium in the Galactic bulge (${l}$ $\approx$ 8.7$^{\circ}$, ${b}$ $\approx$ 2.9, 4.0, 4.7, and 5.7$^{\circ}$). Below galactic
latitudes of 4.0$^{\circ}$, the mid-infrared background spectra in the bulge are similar to the spectra of M and K giants. The UIR emission bands (6.2, 7.7, 8.6, and 
11.3 $\mu$m) are also detected in these regions and likely arise
from the diffuse interstellar medium in the disk. Above galactic latitudes of
4.0$^{\circ}$, the mid-infrared background spectra are similar to the spectra
of those oxygen-rich evolved stars with high mass-loss rates detected by IRAS. One likely
interpretation is that this background emission arises predominantly from those
stars with very low luminosities that have not been detected by IRAS. The
age for such low-luminosity evolved stars could be
15 Gyr, and the existence of a large number of evolved stars with high mass-loss rates in the bulge has a significant impact on our understanding of the stellar content in the Galactic bulge. 
\end{abstract}




%

\section{Introduction}

The Galactic bulge is characterized by a diversity of stellar populations: main-sequence stars, K \& M giants, Asymptotic Giant Branch (AGB) stars, planetary nebulae and white dwarfs. Bulge ages derived from these different stellar populations have big discrepancies. Studies of the RR Lyrae variables, the main-sequence turnoff, the luminosity of M giants, the mass of planetary nebulae, the characteristics of Mira variables, and AGB stars with large mass-loss rates give bulge ages ranging from 5 to 15 Gyr (e.g. Gratton et al. 1986, Terndrup 1988, Frogel \& Whitford 1987, Kinman et al. 1988, Feast \& Whitelock 1987, and van der Veen \& Habing 1990, 1 Gyr = 10$^{9}$ yr). Knowing the stellar populations and their properties in the Galactic bulge would not only provide useful information on the bulge formation, but would also help us to construct stellar synthesis models of spheroidal galaxies, for which only the integrated light can be observed.
From the observations of the Diffuse Infrared Background Experiment (DIRBE) on board the Cosmic Background Explorer (COBE), Arendt et al. (1994) found that the near-infrared background radiation in the Galactic bulge is similar to that of late-K and M giants. Infrared Astronomical Satellite (IRAS) detected a large number of mid-infrared bright stars in the Galactic bulge (Habing et al. 1985), however, the IRAS did not have the sensitivity to detect those mid-infrared faint stars that could dominate the mid-infrared background radiation in the Galactic bulge. Furthermore, the IRAS Sky Survey Atlas (ISSA: Wheelock et al. 1994) flux densities at 12 and 25 $\mu$m cannot provide  unambiguous color data to study the stellar populations, which are likely to be contaminated by the interstellar unidentified infrared (UIR) bands and small grains emission, respectively. 
Thus, spectrophotometric or high spatial-resolution photometric observations in the mid-infrared band (e.g. 4 -- 25 $\mu$m) are required for the study of the mid-infrared background radiation in the Galactic bulge. In this paper we present results from our mid-infrared (4.5 to 11.7 $\mu$m) spectrophotometric observations in parts of the Galactic bulge, and we also discuss the nature of the stellar population there that dominates the mid-infrared background radiation. Throughout our paper we assume a Sun-to-Galactic center distance of 8.5 kpc.  

\section{Observations and Results}

The Mid-Infrared Spectrometer (MIRS) was one of the four science focal-plane instruments that flew aboard the orbiting Infrared Telescope in Space (IRTS). This telescope was a joint Japanese Space Agency (ISAS)/NASA project that was launched on March 18, 1995 and surveyed $\sim$ 7$\%$ of the sky over the course of its 26 day mission life (Murakami et al. 1996). The MIRS had a beam size of \(8'\) $\times$ \(8'\) and operated over a wavelength range of 4.5 to 11.7 $\mu$m with a resolution of 0.23 to 0.36 $\mu$m (Roellig et al. 1994). During the course of its mission, the MIRS observed parts of the Galactic bulge and we present here the spectra at ${l}$ $\approx$ 8.7$^{\circ}$, ${b}$ $\approx$ 2.9, 4.0, 4.7, and 5.7$^{\circ}$. The absolute calibration is about 10$\%$ in this preliminary phase (see Tanab\'e et al. 1996 for details of the MIRS calibration). Selected regions were chosen for this study which did not include bright IRAS point sources and the zodiacal background emission has been subtracted. To subtract the zodiacal background, we first selected zodiacal background spectra that were observed by the MIRS at regions (far away from the galactic plane) that 
have the same ecliptic latitudes but 
different longitudes as those we chose in the bulge. We then corrected the brightness of these observed zodiacal background spectra to match the zodiacal brightness at the same solar elongations as those in the bulge by using a IRAS zodiacal cloud model (Wheelock et al. 1994). Finally, these model zodiacal background spectra subtracted from our selected bulge spectra (bulge + zodiacal emission). The zodiacal cloud emission at 12 $\mu$m in our selected regions 
is between 5 and 9 times higher than the bulge emission there. 

We estimated the extinctions to our observed regions for the purpose of checking whether extinction corrections on our spectra are necessary. Based on the DIRBE/COBE data, Arendt et al. (1994) found that the unreddened near-infrared colors in the Galactic bulge are similar to those of late-K and M giants. Thus, the optical depth at 1.25 $\mu$m can be estimated as

\begin{equation}
    \tau_{1.25\mu m} = \frac{ln[(I_{1.25\mu m}/I_{2.2\mu m})/0.95]}{-0.603}\;,
\end{equation}

where I is the measured DIRBE intensity, 0.95 is the ratio of the flux densities at 1.25 and 2.2 $\mu$m of the unreddened late-K and M giants, and --0.603  = A$_{2.2\mu m}$/A$_{1.25\mu m}$ -- 1 is the reddening between 1.25 and 2.2 $\mu$m and is adapted from the Rieke \& Lebofsky (1985) interstellar extinction law. Based on equation (1) and the final DIRBE/COBE Mission-averaged Pass 3B data with zodiacal light removed as described by Kelsall et al. (1998), the $\tau_{1.25\mu m}$ are calculated by assuming that the extinction lies in front of the emission sources along the line of sight. It should be note that this assumption minimizes the total line-of-sight extinction derived from the reddening. Table 1 summaries the $\tau_{1.25\mu m}$ as derived in the closest DIRBE pixels to our observed positions. The size of a DIRBE pixel is 0.32$^{\circ}$ $\times$ 0.32$^{\circ}$ and the size of DIRBE beam is 0.7$^{\circ}$ $\times$ 0.7$^{\circ}$. With these derived $\tau_{1.25\mu m}$ values, the optical depths at 4.9 $\mu$m ($\tau_{4.9\mu m}$) and 10 $\mu$m silicate ($\tau_{silicate}$) are calculated from the Rieke \& Lebofsky (1985) extinction law and are also summarized in Table 1. In Table 1, the extinctions at/or around our observed positions are small (1 -- 10$\%$ in the MIRS spectral range), therefore, extinction corrections on our spectra were not applied in the following analysis. 

Figure 1 shows the observed spectra of the Galactic bulge at our selected positions. In Figure 1 (a) \& (b), it can be seen that the mid-infrared background spectra in the bulge at Galactic latitudes of 2.9$^{\circ}$ and 4.0$^{\circ}$ are similar to those of M and K giants observed by MIRS, which do not have thick dust shells and the mid-infrared emission mostly arises from the stellar photospheres at temperatures of a few thousands K (Yamamura et al. 1996a,b). The UIR emission bands (6.2, 7.7, 8.6, and 11.3 $\mu$m) are also detected in these regions and they likely arise from the diffuse interstellar medium in the disk (see e.g. Onaka et al. 1996 for the observations of interstellar UIR emission bands in the galactic plane). Figure 2 shows the spectrum of one of the M giants observed by MIRS and it can be seen that its spectral shape is similar to the stellar spectra in Figure 1 (a) \& (b). On the other hand, in Figure 1 (c) \& (d) the mid-infrared background spectra at Galactic latitudes of 4.7$^{\circ}$ and 5.7$^{\circ}$ are similar to those of oxygen-rich AGB stars with thick dust shells, which have a typical silicate absorption feature at 10 $\mu$m (e.g. Forrest et al. 1978 and Herman et al. 1984). For comparison, Figure 3 shows the spectrum of an oxygen-rich AGB star with silicate absorption feature observed by MIRS. Even though the MIRS spectra in Figure 1 (c) \& (d) are a bit noisy the silicate absorption feature is real. This is because the depression in the 8 to 11 $\mu$m range in both of the spectra is $\sim$ --75$\%$ $\pm$25$\%$, which is larger than the 10$\%$ calibration uncertainty of the MIRS. Furthermore, these silicate absorption features should not arise from the line-of-sight extinction, because the optical depths of the interstellar silicate absorption at/or around our observed positions are much less than one (see Table 1). Also shown in Figure 1 are the IRAS Sky Survey Atlas (ISSA) 12 $\mu$m fluxes. In Figure 1 (a) \& (b), the ISSA 12 $\mu$m flux seems to be mostly due to UIR emission bands at 7.7, 8.6, and 11.3 $\mu$m. On the other hand, the ISSA 12 $\mu$m flux in Figure 1 (c) \& (d) is mostly due to circumstellar dust emission. 




\section{Discussion}
\subsection{Evolved stars as seen by IRAS} 

Habing (1987) (hereafter Habing87) has shown that the IRAS 25/12 $\mu$m flux ratio for IRAS point sources in the Galactic bulge can be used to select those late-type stars with high mass-loss rates to produce an estimate of the bulge age. Habing87 assumed that the total integrated flux of these stars can be represented as 3$\nu$F$_{\nu}$(12 $\mu$m), where $\nu$ = 2.5 $\times$ 10$^{13}$ Hz and F$_{\nu}$(12 $\mu$m) is the measured IRAS 12 $\mu$m flux density. The factor 3 is defined as the infrared bolometric correction (see van der Veen \& Breukers 1989). Habing87 proposed that these selected objects in the bulge region are Mira and OH/IR stars, and found that they have a luminosity function which peaks at $\sim$ 4000 L$_{\odot}$. For an AGB star with a luminosity of 4000 L$_{\odot}$, Habing87 estimated that it has an initial mass of 1.7 M$_{\odot}$ and age of $\sim$ 1.3 Gyr (from zero-age main sequence to the first giant branch, Z = 0.04, and Y = 0.3) -- much lower than that estimated from observations in Baade's windows ($\geq$ 10 Gyr) (see e.g. Frogel 1988). Later study by van der Veen \& Habing (1990), with more accurate selection criteria to select AGB stars from the IRAS data, find that the AGB stars with high mass-loss rates in the Galactic bulge have a luminosity function that peaks at $\sim $ 5500 L$_{\odot}$. However, with better knowledge of the evolution of AGB stars, they found that the initial mass of AGB stars with 5500 L$_{\odot}$ is only 1.4 M$_{\odot}$ and their age is 7 -- 15 Gyr (Z = 0.04 and Y = 0.2). Harmon \& Gilmore (1987), however, have found that the AGB stars as seen by IRAS in the bulge have a luminosity function that peaks at $\sim$ 6000 and 7600 L$_{\odot}$. They estimated these AGB stars have initial masses of 1.3 and 1.5 M$_{\odot}$ and ages of 10 and 5 Gyr (Z = 0.1 and Y = 0.25), respectively. 
From the above IRAS studies, the ages of AGB stars in the Galactic bulge have a large range of 5 -- 15 Gyr. The main reason for this large range is due to the high uncertainty of the mass-loss history of these AGB stars in the bulge. 

\subsection{Evolved stars as seen by IRTS}

In Figure 1 (c) \& (d), the mid-infrared background spectra at Galactic latitudes of 4.7$^{\circ}$ and 5.7$^{\circ}$ are similar to the spectra of evolved stars with circumstellar silicate absorption. Therefore, the stars in these regions that produce the mid-infrared background emission are very similar to those AGB stars with large mass-loss rates detected by IRAS. 
Since there are no bright IRAS 12 $\mu$m point sources in these bulge regions that can account for the flux levels measured by the MIRS, one likely interpretation of this background emission is that it arises predominantly from AGB stars with large mass-loss rates, but with very low luminosities that could not have been detected by IRAS. From the IRAS Faint Source and Point Source Catalogs we find that there are less than 50 detected IRAS point sources per square degree at our observed positions in Figure 1 (c) \& (d), which is less than the threshold value (50 per square degree) of IRAS confusion limit at 12 $\mu$m (IRAS Explanatory Supplement: Beichman et al. 1988). Therefore, the IRAS observations at our observed positions in Figure 1 (c) \& (d) are not confusion-limited. Away from confused regions of the sky the sensitivity of IRAS at 12 $\mu$m is 0.5 Jy (Beichman et al. 1988). For an upper-limit estimate of the stellar properties (e.g. luminosity and initial mass), we assume that each of these AGB stars has a 12 $\mu$m flux density of 0.5 Jy, and that this flux density corresponds to a luminosity of 850 L$_{\odot}$ in the bulge. From the MIRS spectra in Figure 1 (c) \& (d), the flux level of 30 Jy at 11.7 $\mu$m implies that there are at least 60 AGB stars in the \(8'\) $\times$ \(8'\) MIRS beam (or $\sim$ 3375 stars deg$^{-2}$), each averaging $\sim$ 850 L$_{\odot}$ (cf. P\'erault et al. 1996 of ISO survey observations of the Galactic plane at ${l}$ $\approx$ --45$^{\circ}$ and ${b}$ $\approx$ 0$^{\circ}$, where they detected $\sim$ 1500 M \& K giants deg$^{-2}$ at 15 $\mu$m). Comparing the spectra in Figure 1, it can be seen that M \& K giants dominate
the mid-infrared emission in the lower latitudes of the bulge (Fig. 1a \& b),
while low luminosity AGB stars with large mass-loss rates dominate the mid-infrared emission 
in the higher latitudes (Fig. 1c \& d). These differences indicate that the bulge
has a gradient in the stellar populations such that the relative
numbers of M \& K giants to AGB stars decrease with increasing latitude.

The relation between the maximum luminosity in the late-stage of AGB star and its initial mass were discussed by Iben \& Renzini (1983), and can be written as

\begin{equation}
 L_{max} \cong 59250 \hspace{0.05in}[ 0.53 \hspace{0.03in}\eta^{-0.082} + 0.15 \hspace{0.03in}(M_{i} - 1) \hspace{0.03in}\eta^{-0.35} - 0.495]\;,
\end{equation}
                         
where L$_{max}$ is the maximum luminosity (in unit of L$_{\odot}$) reached along the AGB evolution, the parameter $\eta$ is proportional to the mass-loss rate of the star and has a value of unity with a factor of 3 uncertainty (1/3 $\leq$ $\eta$ $\leq$ 3), and M$_{i}$ is the initial mass (in units of M$_{\odot}$) of the AGB star. 
Here, we choose $\eta $ = 1 for estimates of the initial mass of our detected AGB stars, and from equation (2) we find that these AGB stars have initial masses of 0.86 M$_{\odot}$. The age of these 0.86 M$_{\odot}$ stars can be estimated if their metallicities are known, unfortunately however, our observations presented here do not provide any useful information about their metallicities. Just for a comparison with the past IRAS studies mentioned in the previous section, we assume that these AGB stars have metallicities of 1.9 times solar (Z = 0.03). From the stellar evolution model of VandenBerg \& Laskarides (1987), the age of a 0.86 M$_{\odot}$ star with Z = 0.03 and Y = 0.35 is $\sim$ 15 Gyr. 

In summary, the major results of this paper are the {following:} (1) there is a 
gradient in the stellar populations of the Galactic bulge, where the relative
numbers of M \& K giants to AGB stars decrease with increasing latitude; 
and (2) there could be a lot of (3375 per square degree) old age ($\sim$ 15 Gyr)
and low luminosity ($\sim$ 850 L$_{\odot}$) AGB stars with large mass-loss rates at the higher latitude field of the bulge. 
  
\acknowledgments
We are grateful to Rick Arendt for providing the optical depths at 1.25 $\mu$m summarized in Table 1 and many useful comments which helped in improving the paper. We thank Martin Cohen for providing the calibrated spectra of standard stars for the calibration of MIRS. We also thank IPAC for their help in the pointing reconstruction of the IRTS and the entire IRTS team for their efforts in ensuring the success of the IRTS mission. I. Y. is supported by the JSPS Research Fellowships for Young Scientists. The COBE data sets were developed by the NASA Goddard Space Flight Center under the guidance of the COBE Science Working Group and were provided by the NSSDC.

\clearpage

\begin{deluxetable}{ccccc}
\footnotesize
\tablecaption{The extinctions in the closest DIRBE pixels to our observed positions. \label{tbl-1}}
\tablewidth{0pt}
\tablehead{
\colhead{galactic coordinates (${l}$ , ${b}$)} & \colhead{(8.78$^{\circ}$ , 2.71$^{\circ}$)}   & \colhead{(8.70$^{\circ}$ , 4.16$^{\circ}$)}   & \colhead{(8.37$^{\circ}$ , 4.72$^{\circ}$)} & 
\colhead{(8.62$^{\circ}$ , 5.60$^{\circ}$)}
} 
\startdata
$\tau_{1.25\mu m}$ &0.415 &0.247 &0.200  &0.122 \nl
$\tau_{4.9\mu m}$ &0.034 &0.020 &0.016 &0.009 \nl
$\tau_{silicate}$ &0.096 &0.057  &0.046  &0.028 \nl
 
\enddata

  
\end{deluxetable}

%
%
%

\clearpage

\clearpage

\figcaption{The observed MIRS spectra at ${l}$ $\approx$ 8.7$^{\circ}$, ${b}$ $\approx$ 2.9, 4.0, 4.7, and 5.7$^{\circ}$. The measured fluxes have been averaged in units of 5 second and this resulted in an effective beam size of \(8'\) $\times$ \(20'\). Also shown in the figure, the $\ast$ symbol, is the IRAS Sky Survey Atlas (ISSA) 12 $\mu$m flux. The bandpass of the IRAS 12 $\mu$m channel was 8 to 15 $\mu$m, and the resolution of the ISSA 12 $\mu$m data is \(4'\). The total flux from the IRAS point sources contributes no more than 10$\%$ of the ISSA 12 $\mu$m flux in these regions. \label{fig1}}

\figcaption{The spectrum of Z UMa (M5 giant) observed by MIRS and adapted from Yamamura et al. (1996b). \label{fig2}}

\figcaption{The spectrum of OH 138.0 +7.2 (an OH/IR star) observed by MIRS. \label{fig3}}

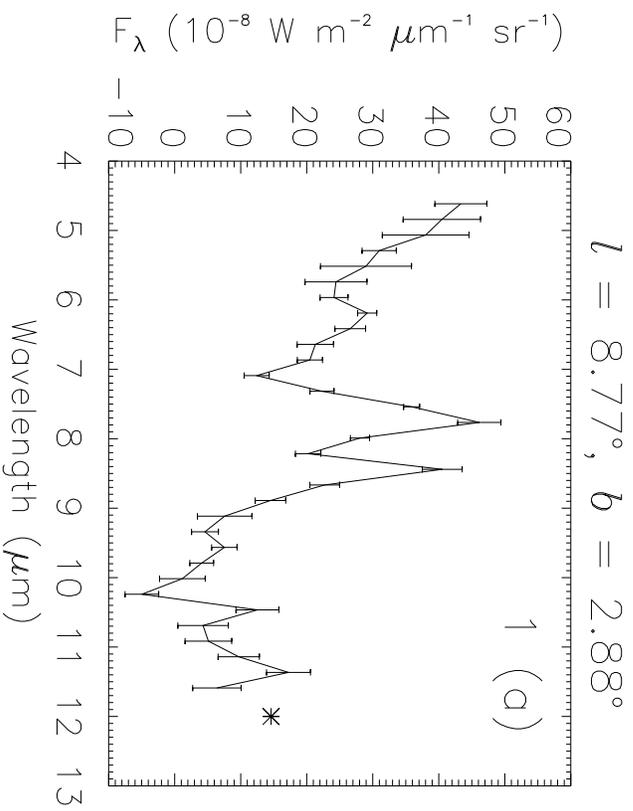
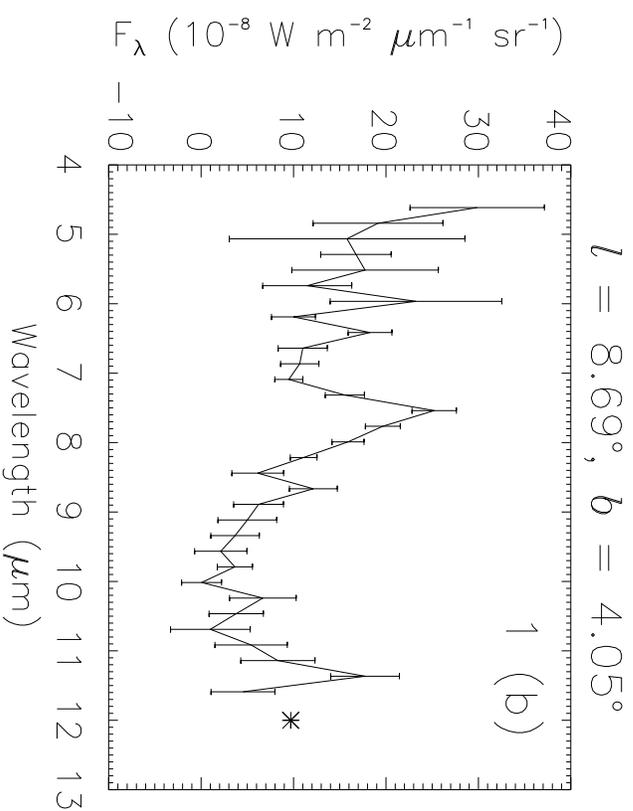
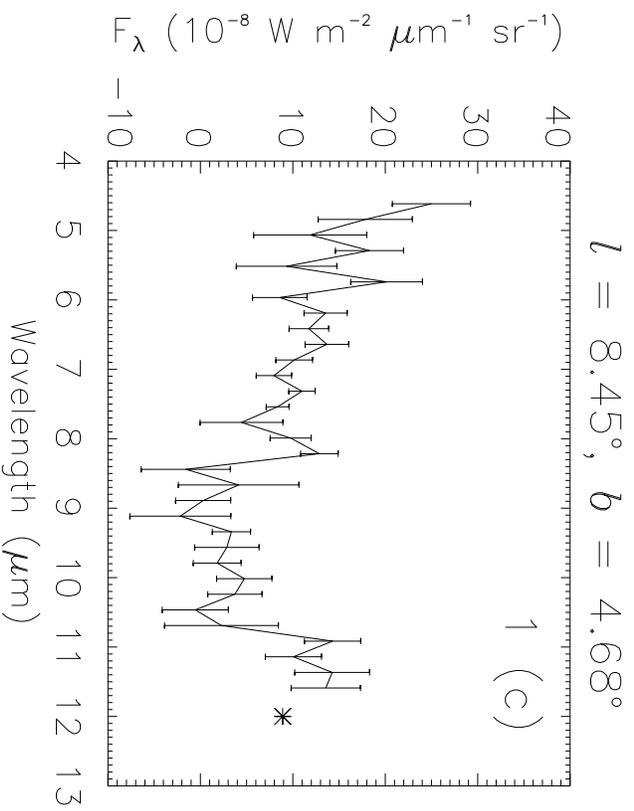
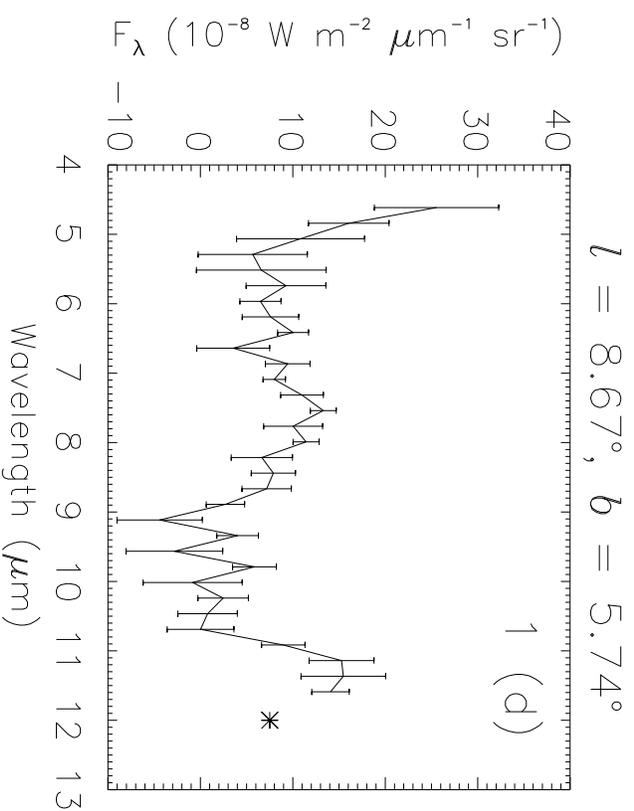
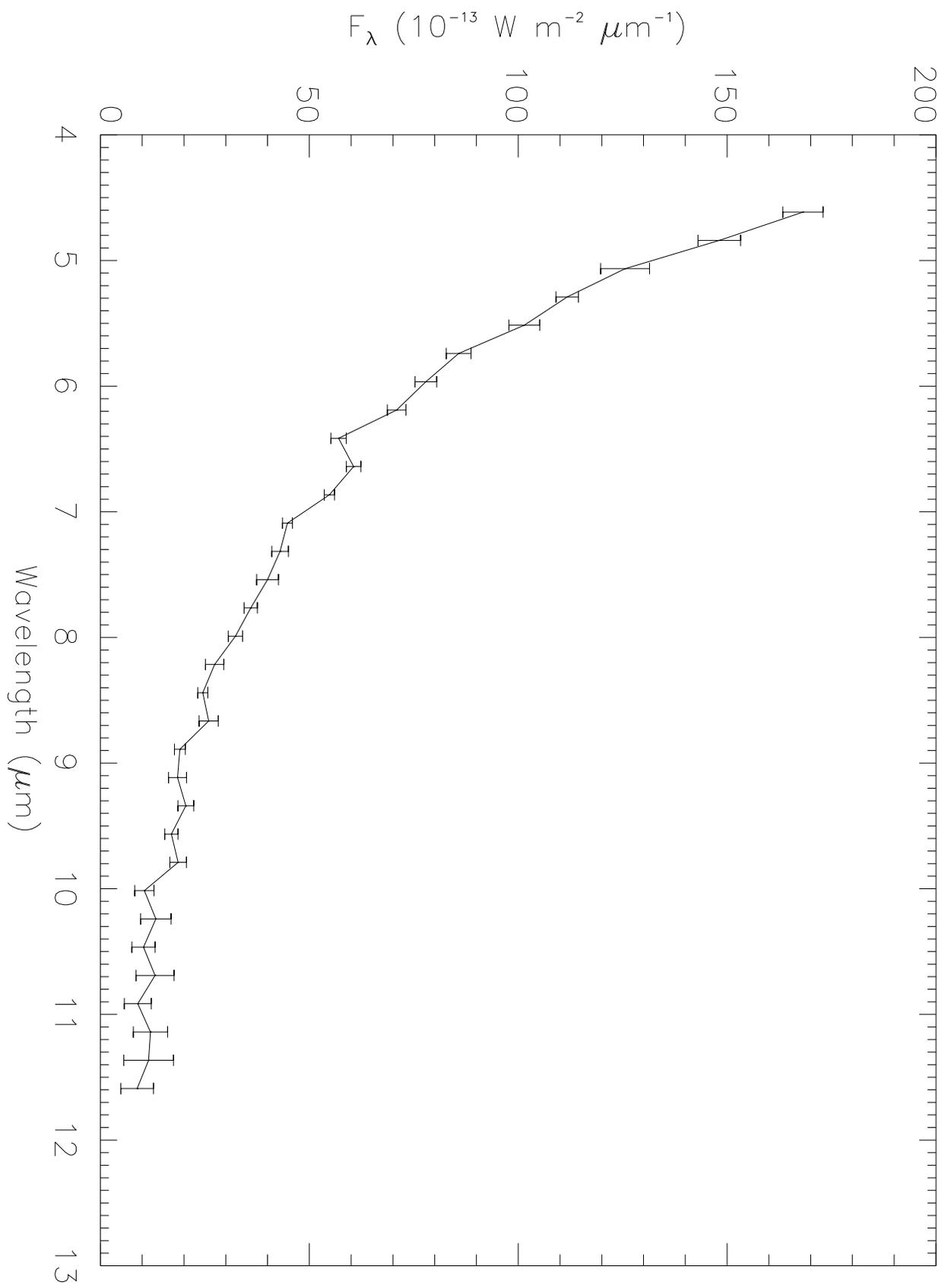
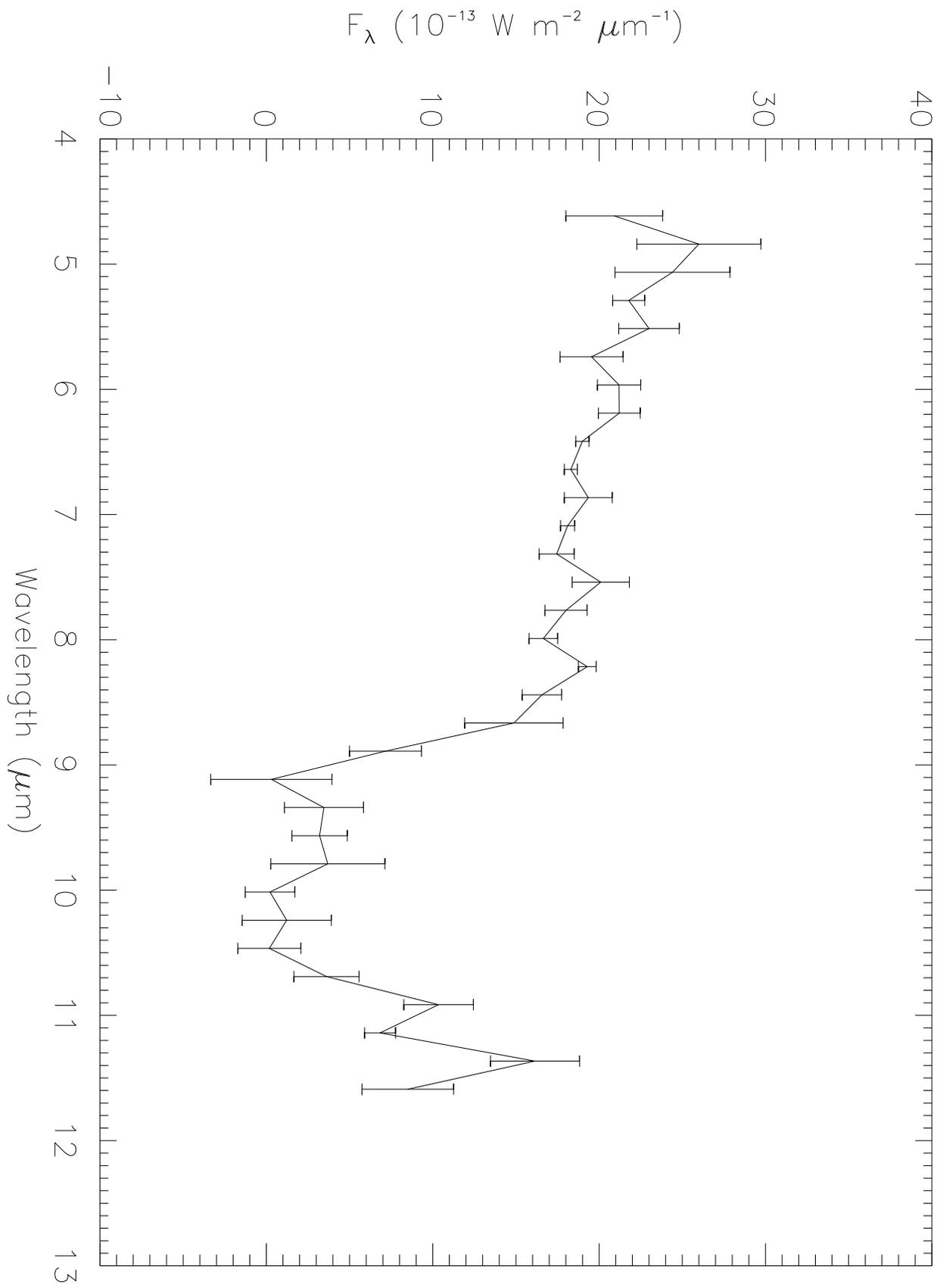
\end{document}